\documentclass[12pt,preprint]{aastex}

\shorttitle{The nature and size of the source in QSO 2237+0305}
\shortauthors{Shalyapin et al.}

\begin{document}

\title{The nature and size of the optical continuum source in QSO 2237+0305}

\author{V.\ N. Shalyapin\altaffilmark{1}, L.\ J. Goicoechea\altaffilmark{2},
D. Alcalde\altaffilmark{3}, E. Mediavilla\altaffilmark{3}, J.\ A.
Mu\~noz\altaffilmark{3}, R. Gil--Merino\altaffilmark{4}}

\altaffiltext{1}{Institute of Radio Astronomy, National Academy of Sciences of
Ukraine, 4, Krasnoznamennaya St., Kharkov, Ukraine, 61002; 
vshal@ira.kharkov.ua}
\altaffiltext{2}{Departamento de F\'{\i}sica Moderna, Universidad de
Cantabria, Avda. de Los Castros s/n, E-39005 Santander, Cantabria, Spain;
goicol@unican.es}
\altaffiltext{3}{Instituto de Astrof\'{\i}sica de Canarias, C/ V\'{\i}a
L\'actea s/n, E-38200 La Laguna, Tenerife, Spain; dalcalde@ll.iac.es,
emg@ll.iac.es, jmunoz@ll.iac.es}
\altaffiltext{4}{Institut fuer Physik, Universitaet Potsdam, Am Neuen Palais 
10, D-14469 Potsdam, Germany; rmerino@astro.physik.uni-potsdam.de}

\begin{abstract}
From the peak of a gravitational microlensing high-magnification event in the A
component of QSO 2237+0305, which was accurately monitored by the GLITP
collaboration, we derived new information on the nature and size of the optical
$V$-band and $R$-band sources in the far quasar. If the microlensing peak is
caused by a microcaustic crossing, we firstly obtained that the standard 
accretion disk is a scenario more reliable/feasible than other usual axially
symmetric models. Moreover, the standard scenario fits both the $V$-band and 
$R$-band observations with reduced chi-square values very close to one. Taking 
into account all these results, a standard accretion disk around a supermassive
black hole is a good candidate to be the optical continuum main source in QSO
2237+0305. Secondly, using the standard source model and a robust upper limit 
on the transverse galactic velocity, we inferred that 90 per cent of the 
$V$-band and $R$-band luminosities are emitted from a region with radial size 
less than 1.2 10$^{-2}$ pc (= 3.7 10$^{16}$ cm, at 2$\sigma$ confidence level). 
\end{abstract}

\keywords{accretion disks --- galaxies: active --- gravitational lensing --- 
quasars: individual (Q2237+0305)}

\section{Introduction}

In the optical continuum, QSO 2237+0305 is a gravitational mirage that 
consists of four compact components (A-D) round the nucleus of the deflector 
(lens galaxy). The light bundles corresponding to the components are passing
through the bulge of the lens galaxy, and thus, if the galactic mass at the 
QSO image positions is mainly in stars, the optical depths to microlensing 
are as high as $\sim 0.5$ (e.g., Schmidt, Webster \& Lewis 1998). In this 
scenario with large normalized surface mass densities, given a component, 
microlensing violent events will result from the source either crossing a 
microcaustic, passing close to a microcusp, or traveling through a network of
microcaustics (e.g., Schneider, Ehlers \& Falco 1992). On the other hand, a 
microlensing violent episode in the light curve of a component can be easily 
distinguished from a intrinsic variation, since the intrinsic variability must 
be observed in all four components of the system with extremely short time 
delays (e.g., Wambsganss \& Paczy{\' n}ski 1994; Chae, Turnshek \& Khersonsky 
1998; Schmidt, Webster \& Lewis 1998). However, when the four brightness 
records of the QSO images have different non-flat shapes, a direct separation 
between the true microlensing fluctuations and the possible intrinsic variation
cannot be achieved. In the case of four incoherent and non-flat observational 
trends, to find the true microlensing behaviours we must do some hypothesis on 
the intrinsic variability. 

Irwin et al. (1989) discovered microlensing variability in the quadruple 
system QSO 2237+0305, and that first evidence was confirmed by other observers
(Corrigan et al. 1991; {\O}stensen et al. 1996; Wo{\' z}niak et al. 2000a,b;
Schmidt et al. 2001; Alcalde et al. 2002). In very recent years, two
gravitational microlensing high-magnification events (HMEs) were clearly
detected by the OGLE collaboration (Wo{\' z}niak et al. 2000b) and 
corroborated by the GLITP monitoring (Alcalde et al. 2002). As each individual
HME is directly related to the intrinsic surface brightness of the source, the 
two HMEs in the $V$ band reported by the OGLE team were used to obtain two 
measurements of both the optical continuum ($V$-band) source size and the rate 
of brightness decline (Shalyapin 2001). Shalyapin (2001) compared the 
observational data included in each HME with the time evolution expected from 
an axially symmetric source crossing a single straight fold caustic. To 
describe the brightness distribution of the $V$-band source, he used a 
power-law model: $I_V(r) = 2^{p_V} I_V (1 + r^2/R_V^2)^{-p_V}$, which is 
determined by a typical radius $R_V$, a typical intensity $I_V = I_V(R_V)$, 
and a power-law index $p_V$. 
The expected microlensing light curves depend on five free parameters, and the 
most relevant ones are $\Delta t = R_V/V_{\perp}$ and $p_V$. We note that a 
direct measurement of the typical radius $R_V$ is not possible, however, using 
some upper limit on the quasar velocity perpendicular to the caustic line 
($V_{\perp}$), it can be obtained a very interesting constraint on the source 
size as measured by means of the typical radius of the 2D brightness 
distribution. A crossing time of $\Delta t \approx$ 90 days is inferred from 
the HME observed in the light curve of the component A, whereas a shorter 
crossing time of about 30 days is consistent with the HME corresponding to the 
image C. However, the whole HME of image A (from day 1200 to day 1800) seems to 
be caused by a complex magnification, which is different to the single straight
fold caustic magnification law. In other words, when the main portion of the 
source is far away from the fold caustic of interest (before day 1400 and after
day 1600), there is evidence for a {\it rare} behaviour, and so, the fit to the
whole microlensing event in the brightness record of the component A could give
biased estimates of the parameters. Taking into account this perspective, we 
only take the results based on the HME corresponding to the image C as 
non-biased parameter estimates. The value of $\Delta t \approx$ 30 days 
together with the velocity constraint by Wyithe, Webster \& Turner (1999) give 
an upper limit on the $V$-band typical radius of $R_V \leq$ 3 10$^{-4}$ pc. 
A similar conclusion was obtained by 
Yonehara (2001), who analyzed the same microlensing event but using a different
picture and a typical microlens mass of $\approx$ 0.1 $M_{\odot}$ (Wyithe, 
Webster \& Turner 2000). On the other hand, Shalyapin (2001) found that the 
value of the power-law index is close to the validity limit of the model ($p_V 
\sim$  1) and the best-fit reduced $\chi^{2}$ is significantly greater than 1. 

Apart from the very recent papers by Shalyapin (2001) and Yonehara (2001),
other previous works also discussed the size of the optical
continuum source (e.g., Wambsganss, Paczynski \& Schneider 1990; Webster et al.
1991; Wyithe et al. 2000). These first studies are based on a poorly sampled
HME which was observed in Q2237+0305A during late 1988. While the continuum
mostly arises from a compact source, the line emission comes from a much larger
region. From two-dimensional spectroscopy of QSO 2237+0305, Mediavilla et al.
(1998) found an arc of extended C {\sc III}]$\lambda$1909 emission that connects 
the components A, D, and B. The observed arc is consistent with a source radius
larger than 100 pc. 

The GLITP (Gravitational Lenses International Time Project) collaboration has
monitored QSO 2237+0305 in the period ranging from 1999 October to 2000
February (Alcalde et al. 2002). The GLITP/PSFphotII photometry in the $V$ and
$R$ bands (see Fig. 1 in Alcalde et al. 2002), showed a peak in the flux of
the component A and a relatively important gradient in the flux of the
component C. These features are related to the two HMEs that were discovered by
the OGLE team. The GLITP light curve for image A traced the peak of the
corresponding HME, i.e., the maximum and its surroundings, with an
unprecedented quality. For example, the OGLE collaboration sampled the $V$-band
peak at 19 dates, whereas the GLITP record included measurements of the
$V$-flux at 52 dates. Moreover, the rest of global behaviours (components B-D)
were accurately drawn from the GLITP photometry (in this paper, we will use the
PSFphotII variant). The $VR$ light curves of the four components A-D have been
also analyzed from a phenomenological point of view, and the global flat shape
for the light curve of Q2237+0305D suggested that the microlensing signal in D
and the intrinsic signal are both globally stationary. In consequence of this
result, it seems that the global variabilities in A-C are unambiguously caused
by microlensing. 

As mentioned here above, the whole HME of Q2237+0305A seems to be originated by
a complex magnification law. However, in principle, the peak of the HME (just 
when the main portion of the source crossed a microcaustic) could be a 
structure mainly caused by a single straight fold caustic, i.e., the curvature 
and other possible close microcaustics do not significantly perturb the simple
magnification law. We adopt this last point of view, and take the GLITP light
curve for image A, including only data points very close to the maximum of the
HME, to be fitted to the microlensing curves resulting from sources crossing a
single straight fold caustic. We remark that the high asymmetry of the peak as 
well as probabilistic arguments are two strong reasons against an 
interpretation of the microlensing peak based on a source passing close to a 
single cusp caustic. 

In Section 2 we present the expected microlensing light curves when axially 
symmetric sources cross a single straight fold caustic, and the fitting
procedure. We use a set of axisymmetric sources: brightness distributions 
enhanced at the centre of the source (standard physical profile, Gaussian 
profile, and $p$ = 3/2, 5/2 power-law profiles) and the uniform brightness 
distribution (e.g., Shakura \& Sunyaev 1973; Schneider \& Weiss 1987; Shalyapin
2001). Section 3 is devoted to the parameter estimation from the comparison 
between the GLITP microlensing peak in the component A and the expected time 
evolutions for the different source models. A discussion on the V-band and 
R-band source sizes, the source size ratio ($R_V/R_R$), and the reliability of 
the source models is also included in Section 3. In this paper, to obtain 
information on the dimension of the $V$-band and $R$-band sources, we will use 
the measurements of the transverse galactic velocity reported by Wyithe, 
Webster \& Turner (1999). Finally, in Section 4 we summarize our results and 
conclusions. 

\section{Theoretical microlensing light curves and fitting procedure}

When a source crosses a single straight fold caustic, a microlensing violent
phenomenon occurs. This is seen as an important fluctuation in the flux of an
image of the source. We are going to describe the properties of a family of
axisymmetric sources, present the time evolution of the flux (microlensing
curve) corresponding to each source model, and finally, introduce the
methodology to compare the theoretical curves with observational data.

\subsection{Source models}

\subsubsection{Uniform and Gaussian disks}

In a given optical band ($U$, $B$, $V$, $R$, $I$, ...), the two simplest 
surface brightness distributions are a uniform disk and a Gaussian disk (e.g., 
Schneider \& Weiss 1987). For the source of uniform intensity, one has a
well-defined source radius $R_{opt}(source)$ that coincides with the typical
radius of the intensity distribution $R_{opt}$. Both radii contain the total 
brightness of the source. However, for the source with Gaussian intensity
distribution and other sources with distributions enhanced at the central
regions, we know a typical radius of the 2D profile ($R_{opt}$), and assume that
$R_{opt}(source) \gg R_{opt}$. This last hypothesis permits us to justify the
approximation $\int_{0}^{R_{opt}(source)} \approx \int_{0}^{\infty}$. For 
example, in the Gaussian case with $I_{opt}(r) = e I_{opt} 
\exp (- r^2/R_{opt}^2)$, the total brightness included in the circle with 
radius $R_{opt}(source)$ is given by $I_{opt}(source) = I_{\infty}\{1 - 
\exp [- R_{opt}^2(source)/R_{opt}^2]\}$, where $I_{\infty}$ is the brightness 
derived from an integration between $r$ = 0 and $r$ = $\infty$. If 
$R_{opt}(source) \gg R_{opt}$, we have $I_{opt}(source) \approx I_{\infty}$. We
note that $I_{opt} = I_{opt}(R_{opt})$ is the intensity at the typical radius 
$R_{opt}$ (i.e., a typical intensity), and so, $I_{opt}(0) = e I_{opt}$ for the
Gaussian law. As different models have typical radii with different meanings, in
a unified scheme, we define two source sizes: the source radius containing half
of the total brightness, $R_{opt}(50\%)$, and the source radius containing 90\%
of the total brightness, $R_{opt}(90\%)$. Using factors $k(50\%)$ and 
$k(90\%)$, which depend on the source model, the radii $R_{opt}(50\%)$ and 
$R_{opt}(90\%)$ can be expressed in terms of the typical radius $R_{opt}$ 
[e.g., $R_{opt}(50\%) = k(50\%) R_{opt}$]. For a uniform disk, the factors are 
$k(50\%) = 1/\sqrt{2}$ and $k(90\%) = 3/\sqrt{10}$, while for a Gaussian disk, 
we obtain $k(50\%) = \sqrt{\ln 2}$ and $k(90\%) = \sqrt{\ln 10}$. 

\subsubsection{Power-law models}

A more interesting set of source models is the family of intensity profiles
with behaviours ($p_{opt} >$ 1)
\begin{equation}
I_{opt}(r) = 2^{p_{opt}} I_{opt} (1 + r^2/R_{opt}^2)^{-p_{opt}} .
\end{equation}
Along with the simplicity of the trends, these power-law models allow to
calculate the expected microlensing curves in an analytical way (see Shalyapin
2001, and the next subsection 2.2.). We shall use two particular values of the
power index, $p_{opt}$ = 3/2, 5/2, which lead to specially simple laws for the
change in flux during a caustic crossing.

For these power-law models, the brightness enclosed within the circle with
radius $r$ will be $I_{opt}(< r) = I_{\infty}[1 - (1 + r^2/R_{opt}^2)^{1 - 
p_{opt}}]$, being $I_{\infty} \approx I_{opt}(source)$. When $r = 
R_{opt}(50\%)$ the second term in the brackets must be equal to 1/2. So 
$k(50\%) = [2^{1/(p_{opt}-1)} - 1]^{1/2}$. On the other hand, it is easy to 
show that $k(90\%) = [10^{1/(p_{opt}-1)} - 1]^{1/2}$.

\subsubsection{Standard accretion disk}

Up to now we described models that are not directly related to physical ideas
about the quasar central engine, i.e., we dealt with effective models. 
However, we can also consider the
standard model of an accretion disk around a supermassive black hole. The
standard Newtonian model (Shakura \& Sunyaev 1973) is based on the supposition
that the released gravitational energy is emitted as a multitemperature blackbody 
radiation. The temperature profile is 
\begin{equation} 
T_s(r) = \left[\frac{3}{8\pi} \frac{GM}{\sigma r^3} \dot{M}
\left(1 - \sqrt{\frac{r_{in}}{r}}\right)\right]^{1/4} ,
\end{equation}
where $\sigma$ is the Stefan constant, $G$ is the gravitation constant, $M$ is
the mass of the central black hole, and $\dot{M}$ is the accretion rate. Here,
$r_{in}$ is the inner radius of the accretion disk, which is usually assumed to
be thrice the Schwarzschild radius of the black hole ($r_{Schw}$). The emitted
intensity obeys a Planck law
\begin{equation} 
I_s(T_s) = \frac{2h\nu_s^3}{c^2} \frac{1}{\exp (h\nu_s/kT_s) - 1}
\end{equation}
and taking into account the redshift of the source $z_s$, the observed
intensity $I = I_s (1 + z_s)^{-3}$ is also a Planck function at the temperature
$T = T_s/(1 + z_s)$,
\begin{equation} 
I(T) = \frac{2h\nu^3}{c^2} \frac{1}{\exp (h\nu/kT) - 1} .
\end{equation}
At a frequency $\nu_{opt}$ (the central frequency of the optical filter), from 
Eqs. (2) and (4) we infer
\begin{equation}
I_{opt}(r) = \frac{(e^w-1)I_{opt}}{\exp \{ (r/R_{opt})^{3/4} \left[ 1 -
u(r/R_{opt})^{-1/2} \right] ^{-1/4} \} - 1} ,
\end{equation}
where $I_{opt} = I_{opt}(R_{opt}) = 2h\nu_{opt}^3/(e^w-1)c^2$, $R_{opt} =
(3GM\dot{M}/8\pi\sigma)^{1/3} [k/h\nu_{opt}(1+z_s)]^{4/3}$, $u = 
(r_{in}/R_{opt})^{1/2}$, and $w = (1 - u)^{-1/4}$. The parameter $u$ is 
taken for simplicity to be negligible, i.e., $u$ = 0 ($w$ = 1), in such a
way that the final intensity profile leads to a theoretical microlensing 
variation quite similar to the exact standard variation at $u <<$ 1. At
$u \sim$ 1 the simplified version of the standard profile is not a useful
approach, but as it will be discussed at the end of Section 3, the observed
light curves do not show the strong {\it break} predicted by an exact model
with $u \sim$ 1 and the simplified profile ($u$ = 0) works reasonably well.  

For the standard source, the coefficients $k(50\%)$ and $k(90\%)$ must be
estimated in a numerical way. Their values are: $k(50\%)$ = 2.386, $k(90\%)$ =
7.038. 

\subsection{Microlensing curves during a caustic crossing}

The microlensing light curves are basically calculated by convolving intensity
distributions with the magnification pattern associated with a single straight
fold caustic. This simple magnification pattern is given by (e.g., Schneider \&
Weiss 1987): a constant background magnification $A_0$ at points located
outside the caustic, and a law $A_0 + a_C/\sqrt{d}$ at points that are placed
inside the caustic. Here, $a_C$ is the caustic strength and $d$ is the
perpendicular distance to the caustic line.

We take a coordinate frame in which the caustic line is defined by the $y$-axis
and the centre of the circular source has the coordinates ($x_c$,0). To obtain
the flux from a part of the disk at ($x$,$y$), its intrinsic intensity should 
be multiplied by both the extinction factor $\epsilon_{opt}$ and the solid angle
$d\Omega$ it subtends on the sky. Considering the relationship: $d\Omega =
A(x,y) d\Omega_*$, where $A(x,y) = A_0 + a_C H(x)/\sqrt{x}$ is the achromatic 
magnification factor, $H(x)$ is the Heaviside step function, and $d\Omega_*$ is
the solid angle in the absence of lens, the elemental flux will be 
$\epsilon_{opt} A(x,y) I_{opt}(r) d\Omega_*$, with 
$r = \sqrt{(x-x_c)^2 + y^2}$. Integrating over the sky, it is inferred a 
radiation flux of the QSO image (e.g., Jaroszy\'nski, Wambsganss \& 
Paczy\'nski 1992)
\begin{equation}
F_{opt}(x_c) = \frac{\epsilon_{opt}}{D_s^2} \int\int A(x,y) I_{opt}(r) dxdy   ,
\end{equation}
where $d\Omega_* = dxdy/D_s^2$ and $D_s$ is the angular diameter distance to the
source. This flux can be rewritten as
\begin{equation}
F_{opt}(x_c) = \frac{\epsilon_{opt} A_0 g_{opt}}{D_s^2} \left[1 + 
\frac{a_C}{A_0}f_{opt}(x_c)\right]   ,
\end{equation}
where $g_{opt} = \int\int I_{opt}(r)dxdy$ represents the total intrinsic 
brightness of the source and
\begin{equation}
f_{opt}(x_c) = \frac{\int\int H(x)x^{-1/2}I_{opt}(r)dxdy}{\int\int 
I_{opt}(r)dxdy} .
\end{equation}
Finally, using normalized coordinates $\xi = x/R_{opt}$ and $\eta = y/R_{opt}$,
one finds $g_{opt} = R_{opt}^2 I_{opt} K$ and $f_{opt}(x_c) = R_{opt}^{-1/2} 
J(x_c/R_{opt})$, being
\begin{equation}
K = \frac{1}{I_{opt}} \int\int I_{opt}[R_{opt}\sqrt{(\xi-x_c/R_{opt})^2 + 
\eta^2}] d\xi d\eta ,
\end{equation}
and
\begin{equation}
J(x_c/R_{opt}) = \frac{\int\int H(\xi)\xi^{-1/2}I_{opt}[R_{opt}
\sqrt{(\xi-x_c/R_{opt})^2 + 
\eta^2}] d\xi d\eta}{\int\int I_{opt}[R_{opt}\sqrt{(\xi-x_c/R_{opt})^2 + 
\eta^2}] d\xi d\eta} .
\end{equation}
The function $J(x_c/R_{opt})$ was also used by Schneider \& Weiss (1987), 
Shalyapin (2001), and other authors.

From the results in the previous paragraph, one obtains
\begin{equation}
F_{opt}(x_c) = F_0 + F_C J(z)   ,
\end{equation}
where $F_0 = (R_{opt}^2/D_s^2)\epsilon_{opt} A_0 I_{opt} K$, $F_C = 
(F_0/\sqrt{R_{opt}})(a_C/A_0)$, and $z = x_c/R_{opt}$. In Eq. (11) two 
chromatic amplitudes appear: the background flux ($F_0$) and the amplitude of 
the contribution due to the extra magnification inside the caustic ($F_C$). On 
the other hand, the constant $K$ and the function $J(z)$ depend on the source 
model, and we focused on the shape factor $J$. In particular we estimated the 
behaviours of $J(z)$ corresponding to the five models discussed in the previous
subsection. For the uniform, Gaussian and standard disks, $J(z)$ was deduced in
a numerical way, whereas for the $p_{opt}$ = 3/2, 5/2 power-law profiles, 
$J(z)$ has an analytical form. In the case $p_{opt}$ = 3/2, Shalyapin (2001) 
gave a simple expression for the shape factor, and when $p_{opt}$ = 5/2,  
\begin{equation}
J(z) = \frac{3\sqrt{1+z^2}-2z}{[2(1+z^2)(\sqrt{1+z^2}-z)]^{3/2}} .
\end{equation}
The trends of $J(z)$ for the five models are depicted in Fig. 1: uniform disk
(dotted line), Gaussian disk (dashed line), $p_{opt}$ = 5/2 power-law model
(dash-dotted line), $p_{opt}$ = 3/2 power-law model (dash-three-dotted line), and
standard accretion disk (solid line). To derive microlensing light curves, the
final step will be to use the trajectory of the centre of the source: $x_c(t) =
V_{\perp}(t - t_0)$, where $V_{\perp}$ is the quasar velocity perpendicular to
the caustic line, and $t_0$ is the time of caustic crossing by the source
centre. We implicitly assumed that the source {\it enters} the caustic, i.e., 
$x_c >$ 0 at $t > t_0$. Moreover, we remark that the time $t$ is measured by
the observer. Inserting the trajectory into Eq. (11), we infer microlensing
curves
\begin{equation}
F_{opt}(t) = F_0 + F_C J\left(\frac{t-t_0}{\Delta t}\right)   ,
\end{equation}
where $\Delta t = R_{opt}/V_{\perp}$ is the crossing time. When the main part 
of an axisymmetric source (i.e., the circle containing half of the total 
brightness) crosses a fold caustic, Eq. (13) will give a good approach to the 
flux of the involved image as a function of time. However, the photometric 
behaviour relatively far from a particular fold caustic could be different to 
the law (13). Firstly, the simple magnification factor $A(x,y) = A_0 + a_C 
H(x)/\sqrt{x}$ may be perturbed as due to the curvature of the fold caustic. 
This problem was studied by Fluke \& Webster (1999), and more recently, by
Gaudi \& Petters (2002). Secondly, the presence of another caustic (i.e., the
existence of a network of caustics) could dramaticly change the magnification
pattern $A(x,y)$ and the predicted time evolution of the flux. Thirdly, the
background magnification $A_0$, which is caused by all the point source images
not associated with the fold of interest, may actually be a function of the
position: $A_0 = A_0(x,y)$. Gaudi \& Petters (2002) introduced microlensing
curves incorporating a slowly varying background. In practice, when one deals
with a given observed microlensing peak or event, the trends (13) must be the
first choices to fit it. Of course a family of reasonable source models should
be tested. After to do the initial fits, if there is evidence for important 
post-fit residues using all the effective and physical scenarios, then some 
corrections must be made (curvature, another caustic, and so on). As we will 
see in Section 3, our dataset does not suggest the need of corrections, and 
the behaviour (13) works very well when some source models are considered. 

\subsection{Theory vs. observations}

Any microlensing peak or event in a component of a lensed quasar can be 
compared with the theoretical light curves presented here above [see Eq. (13)].
Given a source model, the light curve during a caustic crossing (isolated and 
straight fold caustic) depends on 4 parameters:
\begin{enumerate}
\item $F_0$ - a background flux (e.g., in mJy);
\item $F_C$ - a flux related to the extra magnification inside the
caustic (e.g., in mJy);
\item $t_0$ - a time of caustic crossing by the source centre (e.g., in
JD--2450000);
\item $\Delta t$ - a typical crossing time, which is defined as the ratio
between the typical radius $R_{opt}$ and the perpendicular motion $V_{\perp}$ 
(e.g., in days).
\end{enumerate}
So, assuming a particular intensity profile, the task is to estimate the values
of the parameters $F_0$, $F_C$, $t_0$, and $\Delta t$ which best describe the 
observed behaviour. The estimation of model parameters is carried out from a 
fitting method. To fit $N$ observational data $F_{opt}(1),...,F_{opt}(N)$ with 
errors $\sigma_1,...,\sigma_N$, respectively, to the expected ones at times
$t_1,...,t_N$: $F_0 + F_C J[(t_i-t_0)/\Delta t]$, $1 \leq i \leq N$, a 
chi-square minimization will be used. We are going to search for the values of 
the four free parameters ($F_0, F_C, t_0, \Delta t$) which minimize the sum
\begin{equation}
\chi^2(F_0, F_C, t_0, \Delta t)=\sum_{i=1}^N
\left\{\frac{F_{opt}(i) - F_0 - F_C J[(t_i-t_0)/\Delta t]}{\sigma_i}\right\}^2 .
\end{equation}

We must fit a theoretical law that is linear in two parameters ($F_0, F_C$) and
non-linear in $t_0$ and $\Delta t$. On the other hand, to find the best values 
of $F_0$, $F_C$, $t_0$, and $\Delta t$, one must solve the system of equations
$\partial\chi^2/\partial F_0 = \partial\chi^2/\partial F_C = 
\partial\chi^2/\partial t_0 = \partial\chi^2/\partial\Delta t = 0$. As the
function $F_{opt}(t)$ is linear in $F_0$ and $F_C$, the equations 
$\partial\chi^2/\partial F_0 = \partial\chi^2/\partial F_C = 0$ lead to 
analytic relations: $F_0 = F_0(t_0,\Delta t)$ and $F_C = F_C(t_0,\Delta t)$.
From Eq. (14) and these constraints, we can make a new chi-square 
\begin{equation}
\chi^2_*(t_0, \Delta t) = \sum_{i=1}^N
\left[\frac{F_{opt}(i) - F_0(t_0,\Delta t) - F_C(t_0,\Delta t) 
J_i(t_0,\Delta t)}{\sigma_i}\right]^2 .
\end{equation}
It is a clear matter that the function $\chi^2_*$ is the old function 
$\chi^2$ which has been minimized with respect to the two {\it linear} 
parameters, and thus, the complex problem involving 4 free parameters is
reduced to a very simple problem: a minimization of the 2-dimensional 
distribution $\chi^2_*(t_0, \Delta t)$. Assuming we have the best values for 
($F_0, F_C, t_0, \Delta t$), it is necessary to estimate the errors on the 
parameters. In particular, we concentrated on the most relevant parameter 
$\Delta t$. Given the chi-square global minimum $\chi^2 (min) = \chi^2_* 
(min)$, to infer the confidence intervals of a single parameter $\Delta t$, we 
take the different values of the parameter of interest verifying the conditions 
$\Delta \chi^2 = \chi^2_*(t_0, \Delta t) - \chi^2 (min) \leq 1$ (68 percent 
confidence interval, i.e., 1$\sigma$), $\Delta \chi^2 \leq 4$ (95 percent 
confidence interval, i.e., 2$\sigma$), and so on.  
 
To complete the process, we must have an idea of the quality of the fit. If 
the data correspond to the theoretical law and the deviations (due to the
observational noise) are Gaussian, $\chi^2$ should be expected to follow a
chi-square distribution with mean value equal to the degrees of freedom,
$dof = N - 4$. We thus expect $\chi^2$ to be closed to $N - 4$ if the fit is
{\it good}. A quick test is to form the reduced chi-square $\hat{\chi}^2 = 
\chi^2/dof$, which must be close to 1 for a good fit. Moreover, for $dof \geq$
30, the chi-square distribution is essentially normal with standard deviation
of $\sqrt{2\ dof}$. Therefore, the relative deviation $\delta = |\chi^2 -
dof|/\sqrt{2\ dof}$ is expected to be $\leq$ 1.  

\section{Confrontation between GLITP data for Q2237+0305A and theoretical 
light curves}

The results of the comparison between the GLITP $V$-band light curve for
Q2237+0305A and the five expected time evolutions are presented in Table 1, 
while the results from the comparison in the R band appear in Table 2. We are 
mainly interested in measurements of the $V$-band and $R$-band source sizes, 
and so, only the uncertainties (1$\sigma$ intervals) in the estimates of the 
relevant parameter $\Delta t$ are quoted. In Table 3 (second column) we have
included the 1$\sigma$ confidence intervals of the source size ratio ($q = 
R_V/R_R$). Table 3 also includes upper limits on the half-light radii. As 
$R_{opt}(50\%) = k(50\%) V_{\perp} \Delta t$, to infer the constraints, we used
$V_{\perp} <$ 7000 km s$^{-1}$ (Wyithe, Webster \& Turner 1999) and $\Delta t 
\leq \Delta t_{max} = \Delta t_{best-fit} + \sigma_+(\Delta t)$. For some 
source models, information on the source radii (containing 100\% of the total 
brightness) or $R_{opt}(90\%)$ in the $V$ and $R$ bands, is quoted for 
discussion. Wyithe, Webster \& Turner (1999) determined upper limits on the 
transverse galactic velocity ($v_t$) from the distribution of light-curve 
derivatives. In Fig. 12 (bottom panels) of Wyithe, Webster \& Turner, it was 
presented the 95 per cent upper limit to $v_t$ as a function of the adopted 
random errors, the microlens mass distribution, and the source details. For 
mass distributions which are characterized by a mean microlens mass equal to 
or less than 1 $M_{\odot}$, and any reasonable choice of the random 
uncertainties, the global bound is of $v_t \leq$ 900 km s$^{-1}$. This bound is
valid for any source size and intensity profile, and we assumed it as an
absolute upper limit. Using the relationship $V = (D_s/D_d) v_t$, where $D_d$ 
is the angular diameter distance to the lens (deflector) and $V$ is the 
transverse quasar velocity, one easily derives that $V_{\perp} < V \leq$ 7000 
km s$^{-1}$ ($\Omega$ = 1). 

The upper limits on the $V$-band and $R$-band source sizes are presented in 
Table 3. For the uniform disk: $R_V, R_R <$ 6.3 10$^{-4}$ pc (= 1.9 10$^{15}$ cm). 
For the $p$ = 3/2 power-law 
model and the standard model, the radii containing 50\% of the total disk 
brightness may be as large as 2.7 10$^{-3}$ pc, and interestingly enough the 
radii containing almost all the brightness (90\%) must be smaller than 1.2 
10$^{-2}$ pc (= 3.7 10$^{16}$ cm), in a reasonable agreement with the expected 
radial size of a hot accretion disk around a 10$^8$ $M_{\odot}$ black hole 
(10$^3$ $r_{Schw} \approx$ 10$^{-2}$ pc). With respect to the parameter $q$, a 
value of $q$ less than 1 would suggest the existence of two concentric sources 
with different radii, the $V$-band source being inside the $R$-band one. 
Unfortunately, although the preferred values of $q$ vary in the interval 
0.65 -- 0.98, i.e., $q <$ 1, the error bars are quite large and we cannot
fairly distinguish between the cases $q <$ 1 and $q$ = 1. Only the Gaussian 
model led to $q \leq$ 0.86 at 1$\sigma$ confidence level. For the standard 
disk, $R_{opt} \propto \nu_{opt}^{-4/3}$ and $q = (\nu_R/\nu_V)^{4/3} \approx$ 
0.8. However, this standard source size ratio cannot be confirmed from our 
indirect measurement. 

In Table 1, the $\delta (min)$ values are not good for the $V$-band uniform and
Gaussian disks (see subsection 2.3). However, it is evident that both the 
$p_V$ = 3/2 power-law model and the $V$-band standard disk work very well. The 
intensity distribution with a power-index of 3/2 and the standard accretion 
disk are also favored from data in the $R$ band (see Table 2). For the standard 
source, curves of constant chi-square in the $t_0$ (JD--2450000)--$\Delta t$ 
(days) plane are showed in Fig. 2. In Fig. 2 we can see the contours 
associated with $\Delta \chi^2$ = 1 and $\Delta \chi^2$ = 4. On the other hand, 
in Fig. 3 we drawn together the observed light curves and the corresponding 
{\it standard} fits. The agreement between GLITP observations and fits is 
excellent, and taking as reference the day 1500, it is evident the existence 
of an important asymmetry. For comparison, in the top panel (dashed line), we
also drawn the best-fit from the theoretical microlensing curve related to the
exact standard accretion disk (including realistic edges). As the exact
microlensing law fits the observations slightly better than the approximated one,
we confirm the high feasibility of the model based on physical grounds. However, 
although the standard disk led to values of $\hat{\chi}^2 (min) \approx$ 1, the
$p$ = 3/2 power-law model is also in good agreement with the observations. As a
consequence of this result, we tested if the $p$ = 3/2 power-law model is a
simplified version of the exact standard model or, on the contrary, we are 
handling two very different scenarios. The most simple test is a comparison of the
1D intensity profiles, i.e., the integrals of the 2D intensity distributions along 
the direction defined by the caustic line. Therefore we first simulated 
a 2D intensity distribution associated with a standard accretion disk (exact model). 
Then they were derived the standard 1D profile and all the effective 1D profiles 
fitting the standard one. The comparison between the standard behaviour and the 
effective trends appears in Fig. 4. The filled circles represent the standard 1D 
profile (including the two typical "horns" close to $x$ = 0), while the solid lines 
trace the power-law ($p$ = 3/2) 1D profile (left-hand top panel), the power-law ($p$ 
= 5/2) 1D profile (right-hand top panel), the Gaussian 1D profile (left-hand bottom 
panel), and the uniform 1D profile (right-hand bottom panel). It is now clear that 
the $p$ = 3/2 power-law model very closely mimics the behaviour of the standard 
scenario, i.e., it is a rough version of the exact standard model.

\section{Conclusions}

We have analyzed the peak of a microlensing high-magnification event which was
accurately monitored by the GLITP collaboration (Alcalde et al. 2002). The
prominent event has occurred in image A of the quadruple system QSO 2237+0305 
(Wo{\' z}niak et al. 2000b), and the GLITP team observed its peak in two
optical bands ($V$ and $R$). As both $V$-band and $R$-band light curves are
characterized by high flux and time resolutions, we attempted to interpret the
observational trends in terms of some theoretical microlensing light curves. In
principle, the observed microlensing peak could mainly result from the source
crossing a microcaustic or passing close to a microcusp. However, taking into
account probabilistic arguments and the important asymmetry around day 1500
(see Fig. 3), we have discarded the hypothesis of a source in the vicinity of
a single cusp caustic. Thus we only considered the most simple alternative
picture: a source crossing a single straight fold caustic. Different axially
symmetric source models were used, and consequently, several theoretical
microlensing curves were compared with the observed ones. To discuss the
reliability/feasibility of different intrinsic intensity profiles, all source
models were chosen to cause theoretical microlensing curves with the same number 
of free parameters. These parameters are two characteristic fluxes, the time of
caustic crossing by the source centre, and the typical crossing time.

Our main results and conclusions are:
\begin{enumerate}
\item From the fits and the upper limits on the transverse galactic velocity
claimed by Wyithe, Webster \& Turner (1999), we inferred constraints on the 
$V$-band and $R$-band source sizes. For a uniform source model, the $V$-band 
and $R$-band radii should be less than 6.3 
10$^{-4}$ pc = 1.9 10$^{15}$ cm, at 1$\sigma$ confidence level. For a Gaussian
disk, the typical $V$-band radius is $<$ 7.8 10$^{-4}$ pc (= 2.4 10$^{15}$ cm, 
at 1$\sigma$ CL), while the typical $R$-band radius can be a little larger ($<$
1.6 10$^{-3}$ pc = 4.9 10$^{15}$ cm, at 1$\sigma$ CL). Taking the total size of
the $R$-band source as the source diameter for a uniform disk, or the
full-width at one-tenth maximum for a Gaussian profile, we found that
the 1$\sigma$ upper limits on the $R$-band source size are 3.9 10$^{15}$ cm
(top-hat) and 1.5 10$^{16}$ cm (Gaussian). These bounds agree well with the
results by Wyithe et al. (2000). On the other hand, for 
the standard accretion disk, we obtained that 90\% of the $V$-band and $R$-band
luminosities are radiated within a radius of 8 10$^{-3}$ pc (= 2.5 10$^{16}$ 
cm, at 1$\sigma$ CL), 1.2 10$^{-2}$ pc (= 3.7 10$^{16}$ cm, at 2$\sigma$ CL). 
As the total luminosity of a standard accretion disk around a 10$^8$ 
$M_{\odot}$ black hole will be probably enclosed in a circle with radial size
of $\approx$ 10$^{-2}$ pc, the {\it standard} results are highly consistent
with the current paradigm on the central engine in QSOs. However, we remark 
that other central masses also agree with the constraints. Once upper limits
of $\sim$ 10$^{15}$ -- 10$^{16}$ cm are known, we may attempt to test the
hypothesis of a negligible caustic curvature (at $V_{\parallel} <$ 7000 
km s$^{-1}$, the parallel path length during $\sim$ 100 days will be also
less than 10$^{15}$ -- 10$^{16}$ cm). The typical caustic curvature radius
is usually assumed to be the Einstein-ring radius on the source plane, i.e., 
$R_{CC} \sim R_E \sim 10^{17}(m/M_{\odot})^{1/2}$ cm, where $m$ is the
microlens mass ($\Omega$ = 1, $H$ = 60 -- 70 km s$^{-1}$ Mpc$^{-1}$). 
Therefore, for $m \sim 1 M_{\odot}$, the ratios between the optical radii
and the typical caustic curvature radius will be smaller than 0.01 -- 0.1,
and this result supports the straight fold caustic approximation. For a
small microlens mass ($m \sim 0.1 M_{\odot}$), we however have a smaller
$R_E$ and cannot confirm the weakness of the curvature effects.   
\item We also studied the source size ratio, i.e., the ratio between the
$V$-band radius and the $R$-band radius. The source size ratio values are 
independent of the transverse quasar velocity. At 1$\sigma$ CL, only the
Gaussian source model led to a $V$-band source being inside the $R$-band one.
For the standard source, one hopes for a ratio of about 0.8, but this expected
value could not be confirmed from our microlensing experiment. The 1$\sigma$
confidence interval (0.53 -- 1.26) is in agreement with the two possible
situations: the $V$-band source being inside the $R$-band source, and both
sources having a similar size.
\item An important issue is the reliability/feasibility of several source 
models tested by us. With respect to this point, we deduced very interesting
conclusions. The usual top-hat and Gaussian profiles are not favored from the
data in the $V$ and $R$ bands. The results are better when power-law profiles
are assumed, particularly for the model with a power index of 1.5, which more
closely resembles the one-dimensional intensity profile of the exact standard 
accretion 
disk. For the standard source model, the reduced chi-square values are very 
close to one. So, an accretion disk around a supermassive black hole seems a good 
candidate to be the optical continuum main source in QSO 2237+0305 (see Fig. 3), 
and a measurement of the black hole mass and the mass accretion rate could be 
made in that system (e.g., Yonehara et al. 1998). This last topic will be 
discussed in a separate paper. We also note that a hybrid scenario in which the
light in the $V$ band is emitted from an accretion disk and the $R$-band light
comes from the accretion disk and another extended region, is not in contradiction
with the observed microlensing peaks. This result totally agrees with the main
conclusion of Jaroszy\'nski, Wambsganss \& Paczy\'nski (1992), who previously
tested the hybrid model from old observational data. The assumption of a 
complementary $R$-band extended source only modifies the expression of the
$R$-band background flux $F_{0R}$ in Eqs. (11) and (13), and thus, the 
interpretation of its best-value and uncertainties. Exclusively using GLITP
data, one can try to analyze the possible existence of a light excess in the
$R$ band. To do this task, it is needful an accurate calibration of the flux
in both optical bands and small uncertainties in the measurements of the 
background fluxes. A detailed study of this topic is however out of the scope 
of the paper, which is focused on the optical continuum main (compact) source.
On the other hand, in order to fit the observed spectrum of the quasar, Rauch
\& Blandford (1991) did not consider additional $R$ light coming from a large
region, but adopted a non-classical accretion disk model. Unfortunately, the
disk model by Rauch \& Blandford (1991) cannot account for the old microlensing
variations in QSO 2237+0305.
\end{enumerate}

We remark that the OGLE collaboration also monitored the $V$-band
event in Q2237+0305A 
(http://www.astro.princeton.edu/$\sim$ogle/ogle2/huchra.html). The GLITP
$V$-band photometry traced the peak of the microlensing event, whereas the OGLE
$V$-band dataset described the behaviour of the whole fluctuation. In
comparison with the OGLE observational procedure, the GLITP observations are of
higher quality because they were obtained with a larger telescope, using a
detector with better resolution, and on nights with better seeing. Therefore,
as due to the quality of the observations and the excellent sampling rate, the
GLITP $V$-band peak is probably the best tracer of the underlying signal around
the maximum of the whole flux variation. On the other hand, the whole event
seems to be caused by a complex magnification, which may include ingredients
such as the curvature of the fold caustic (e.g., Fluke \& Webster 1999), the
presence of another caustic, and a non-constant background magnification (e.g.,
Gaudi \& Peters 2002). The simple fit to the whole event may thus give a wrong
estimation of the parameters. In any case, we chose two OGLE observation
periods to infer {\it standard} solutions and compare them with the {\it 
standard} fit presented in Table 1. The first period included only the end of
1999, from day 1450.6 to day 1529.5. The dataset is called OGLE99, and it 
corresponds to the GLITP monitoring period. The second dataset (OGLE99-00)
covers the 1999-2000 seasons, more exactly from day 1289.9 to day 1766.7. In
Table 4 we can see all at once the fits from the GLITP, OGLE99, and OGLE99-00 
light curves. As $\hat{\chi}^2 (min) >$ 1 for the OGLE best solutions, in the
time parameter estimation from the OGLE datasets, we considered the 1$\sigma$ 
bounds associated with $\Delta \chi^2 = \hat{\chi}^2 (min)$ rather than 
$\Delta \chi^2$ = 1 (e.g., Grogin \& Narayan 1996). The GLITP and OGLE99 fits 
are not consistent each other, but the amplitudes and the crossing time from 
the OGLE99-00 brightness record are close to the values of $F_0$, $F_C$, and 
$\Delta t$ from the GLITP dataset.

Finally, we must emphasize that an important progress can be made from 
accurate and detailed data of a microlensing peak. One can discuss on the
reliability/feasibility of different source models leading to the same
number of free parameters, and thus, to discard some of them and to find
the models that agree with the observations. Given a good model, which is
consistent with the data, it is possible to obtain a robust upper limit on the
size of the optical source. If, for example, the good model is an accretion 
disk around a massive black hole, then one can try to work out a technique to
measure the central mass and the accretion rate. Moreover, as the accurate
and well-sampled microlensing peaks seem to be inconsistent with some intensity
profiles, it could be very interesting to apply the deconvolution method to
these peaks and to derive the best brightness distribution in a more direct way
(Grieger, Kayser \& Schramm 1991; Agol \& Krolik 1999; Mineshige \& Yonehara 
1999). Although we had not success in the accurate and robust indirect estimation 
of the ratio between the $V$-band radius and the $R$-band radius (the parameter 
$q$), new multiband monitoring campaigns could also lead to relevant measurements 
of source size ratios. With regard to this last issue, we note that a direct 
estimate of $q$ from the cross-correlation of our $V$-band and $R$-band light 
curves is now in progress, but the expectations are not very promising.

\acknowledgments

We would like to thank Joachim Wambsganss for comments on a first version of 
the paper. We also acknowledge the anonymous referee for some interesting 
suggestions. The GLITP observations were made with Nordic Optical Telescope 
(NOT), which is operated on the island of La Palma jointly by Denmark, Finland,
Iceland, Norway, and Sweden, in the Spanish Observatorio del Roque de Los 
Muchachos of the Instituto de Astrof\'{\i}sica de Canarias (IAC). We are 
grateful to the technical team of the telescope and E. Puga for fulfill almost 
all the observational work. This work was supported by the P6/88 project of the
IAC, Universidad de Cantabria funds, DGESIC (Spain) grant PB97-0220-C02, and 
the Spanish Department of Science and Technology grants AYA2000-2111-E and 
AYA2001-1647-C02.

\clearpage

{\bf FIGURE CAPTIONS}

\figcaption[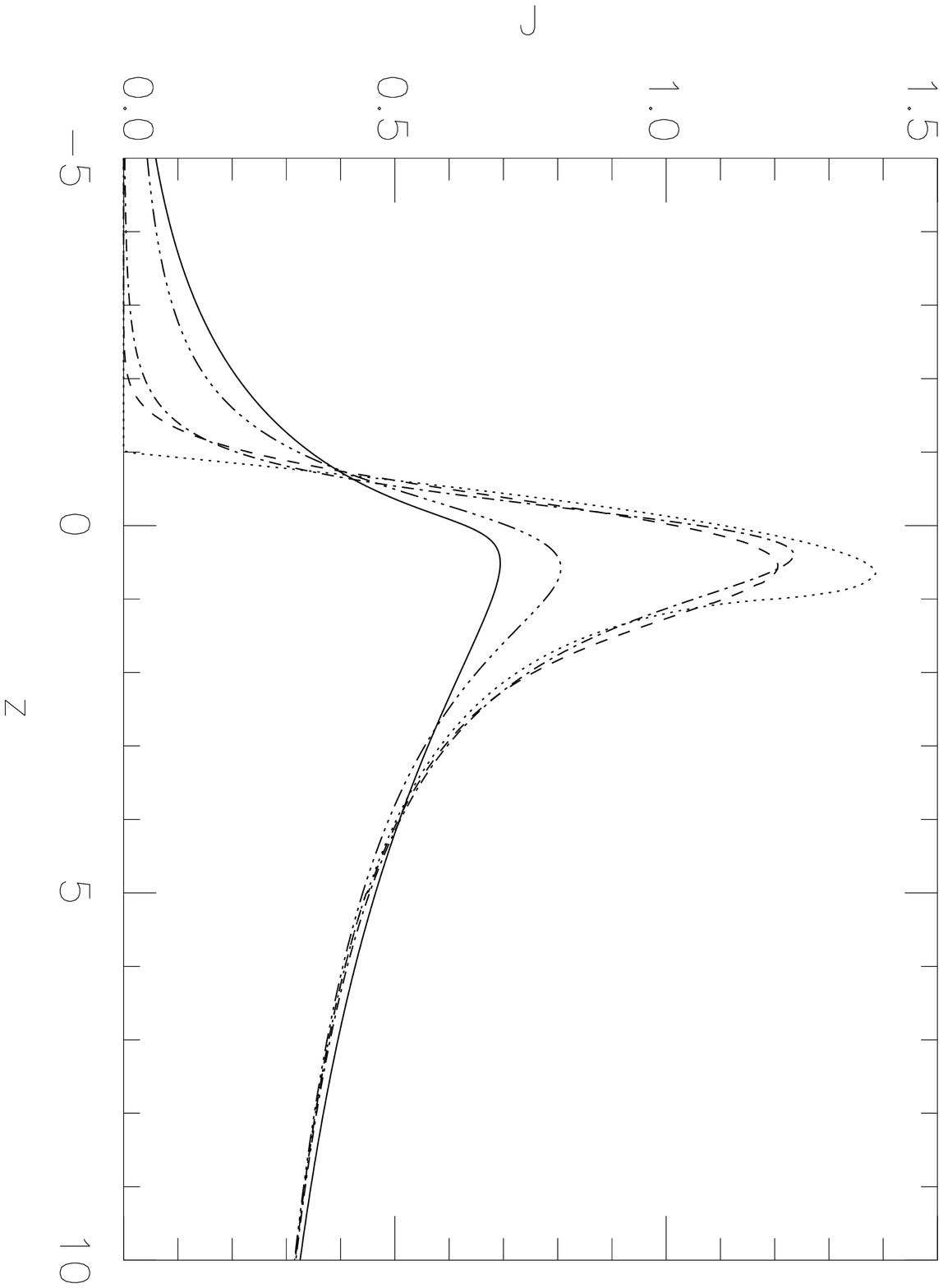]{Shape factor $J(z)$. Five source models are considered:
uniform disk (dotted line), Gaussian disk (dashed line), $p_{opt}$ = 5/2 
power-law model (dash-dotted line), $p_{opt}$ = 3/2 power-law model 
(dash-three-dotted line), and standard accretion disk (solid line).}

\figcaption[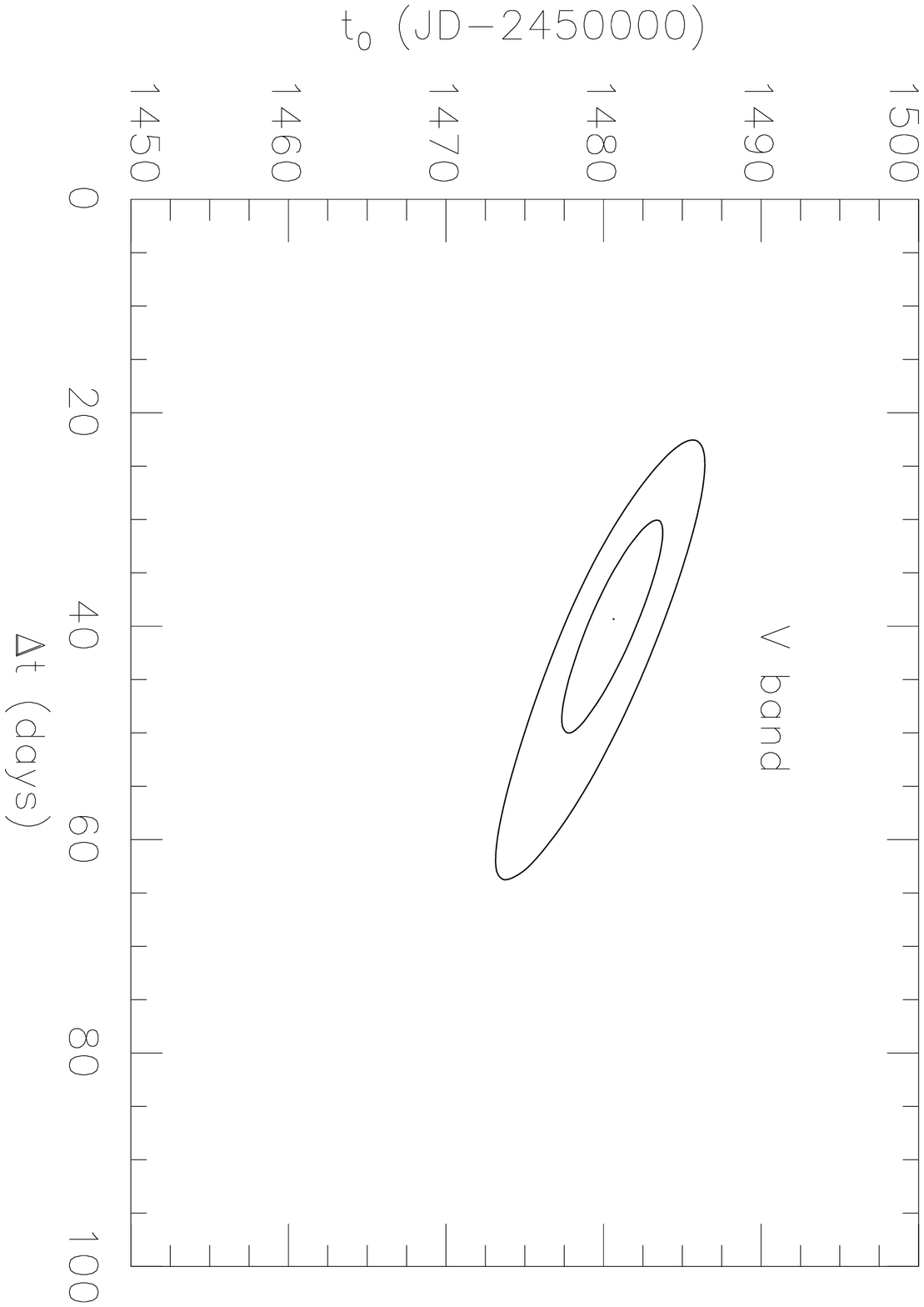,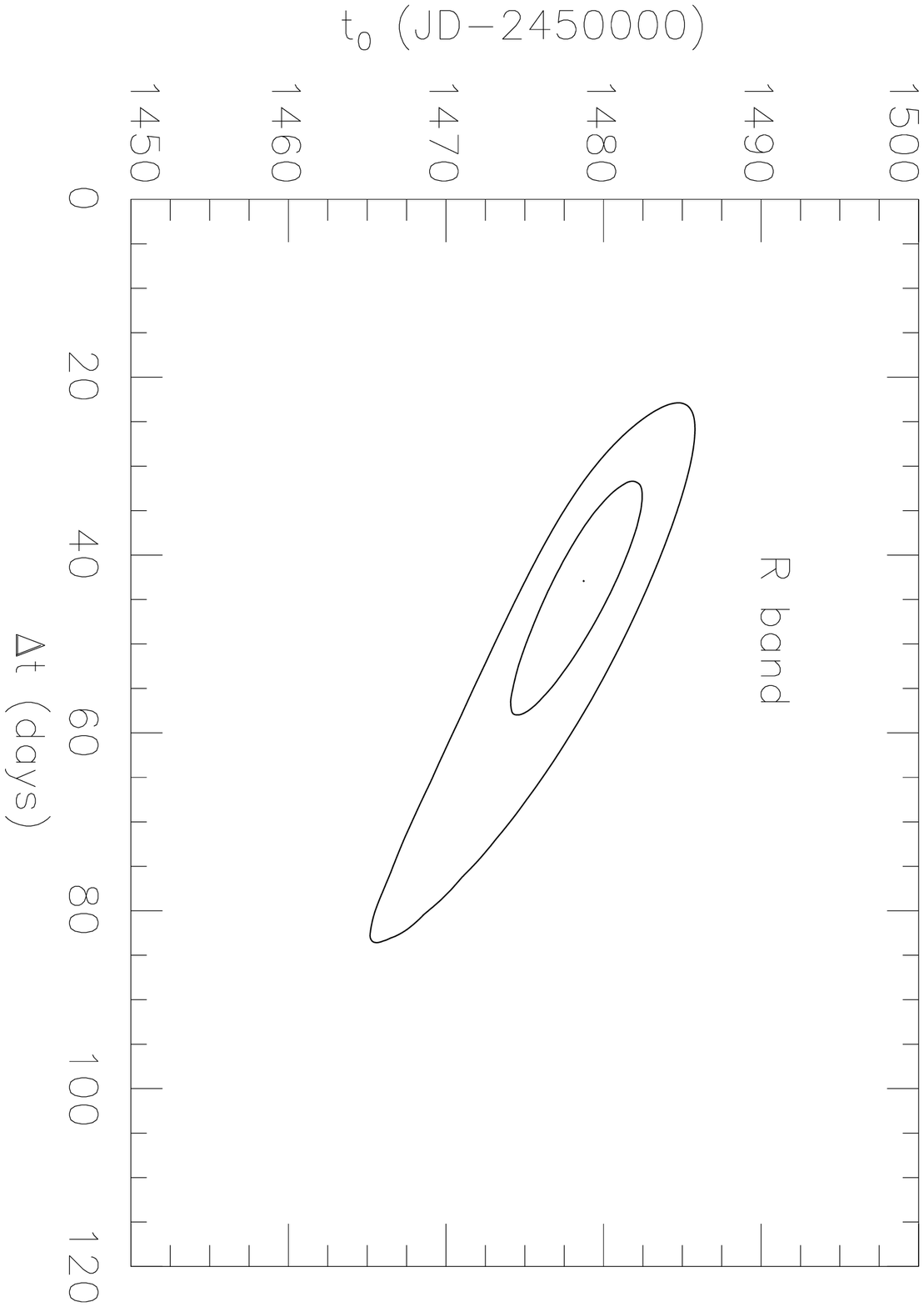]{$\Delta \chi^2$ = 1 (interior solid lines) 
and $\Delta \chi^2$ = 4 (exterior solid lines) contours in the $t_0$ 
(JD--2450000)--$\Delta t$ (days) plane. The source is assumed to be a standard 
accretion disk. The datasets are: the GLITP $V$-band light curve of Q2237+0305A 
(top), and the GLITP $R$-band light curve of Q2237+0305A (bottom).}

\figcaption[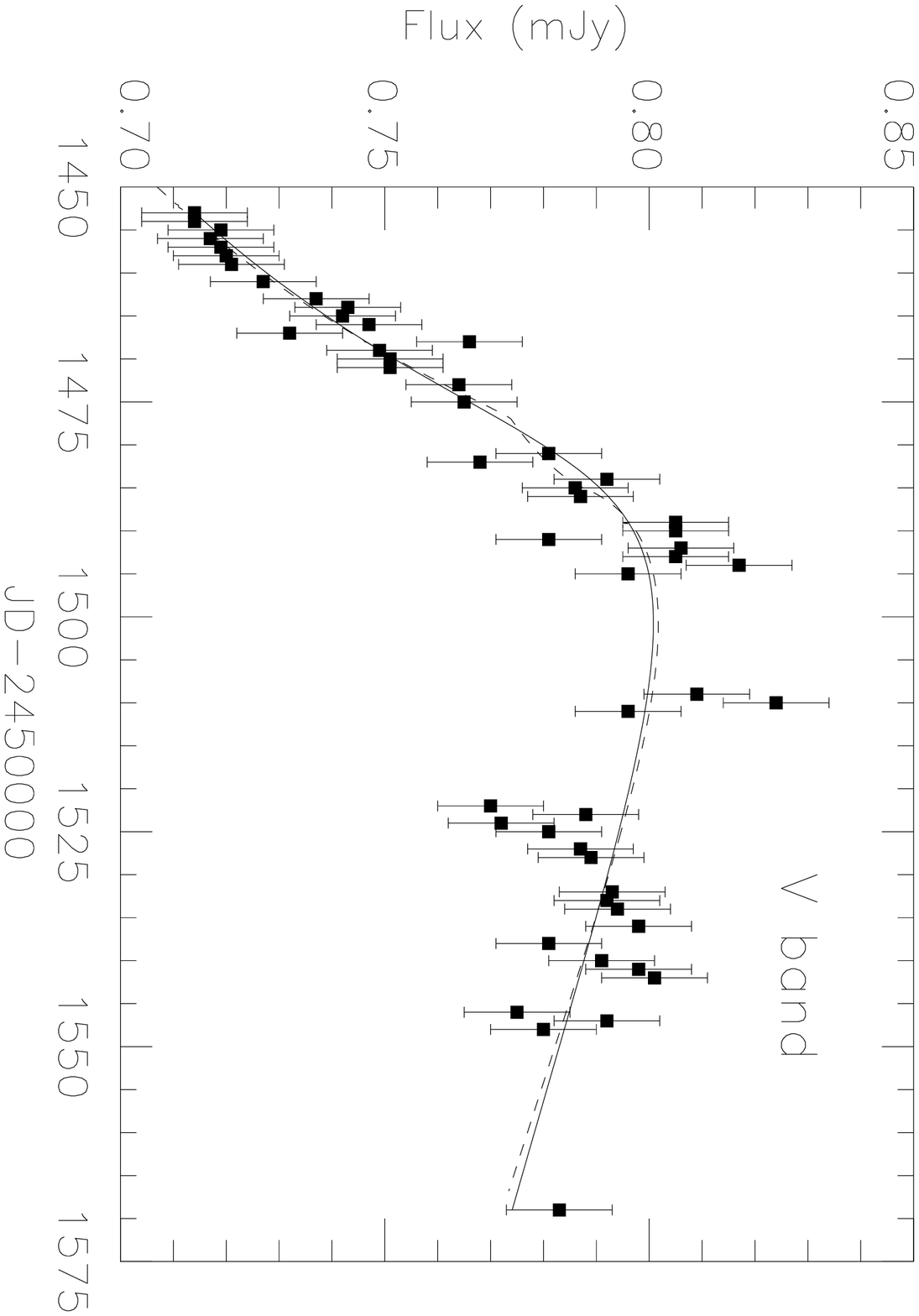,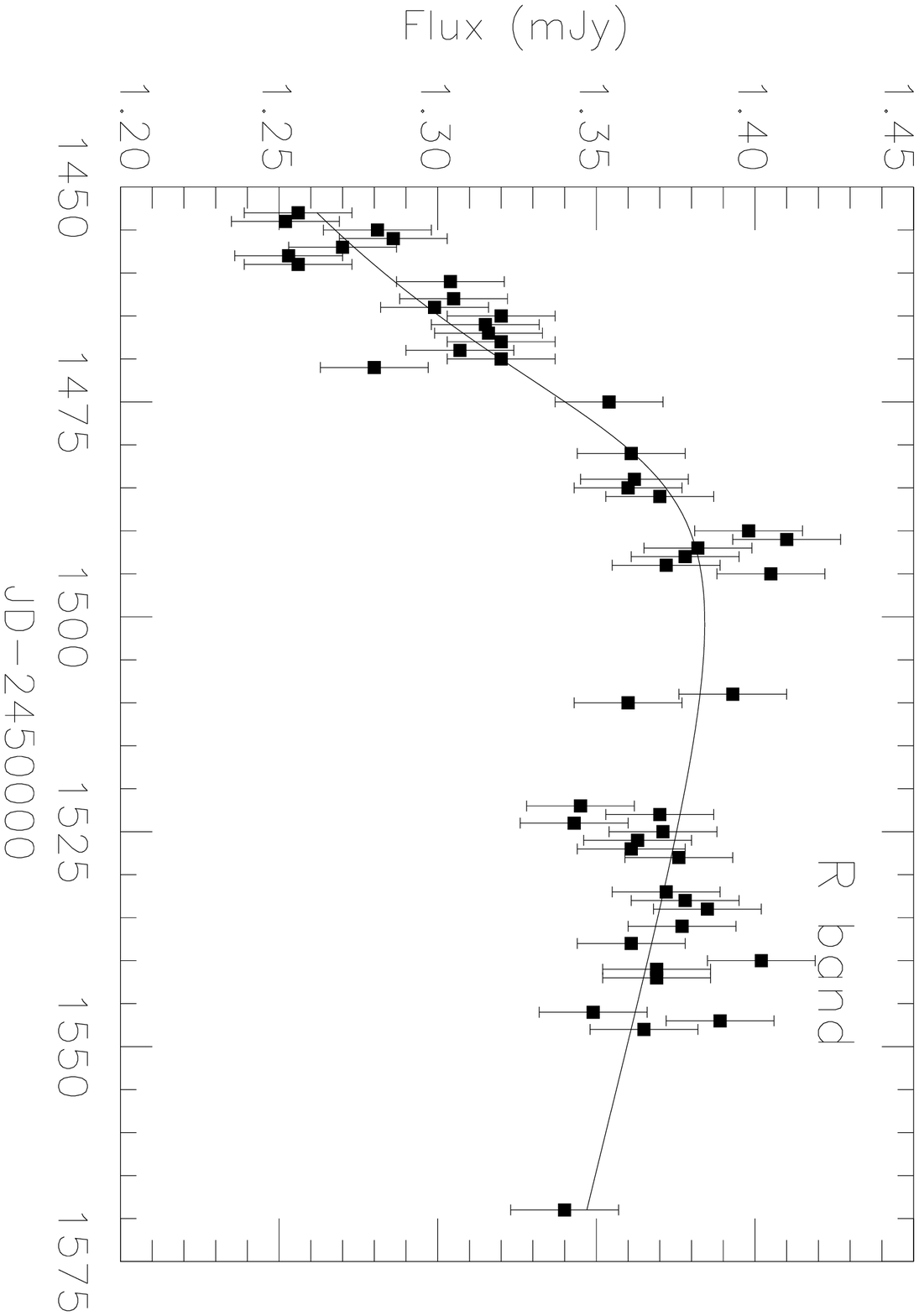]{Observed light curves and the corresponding 
{\it standard} fits. Top: $V$-band. Bottom: $R$-band. The dashed line in the top
panel ($V$-band) represents the best-fit from an exact standard accretion disk
crossing a caustic line, and the weak {\it break} around day 1480 is an edge 
effect. We can see a good agreement between both $V$-band fits.}

\figcaption[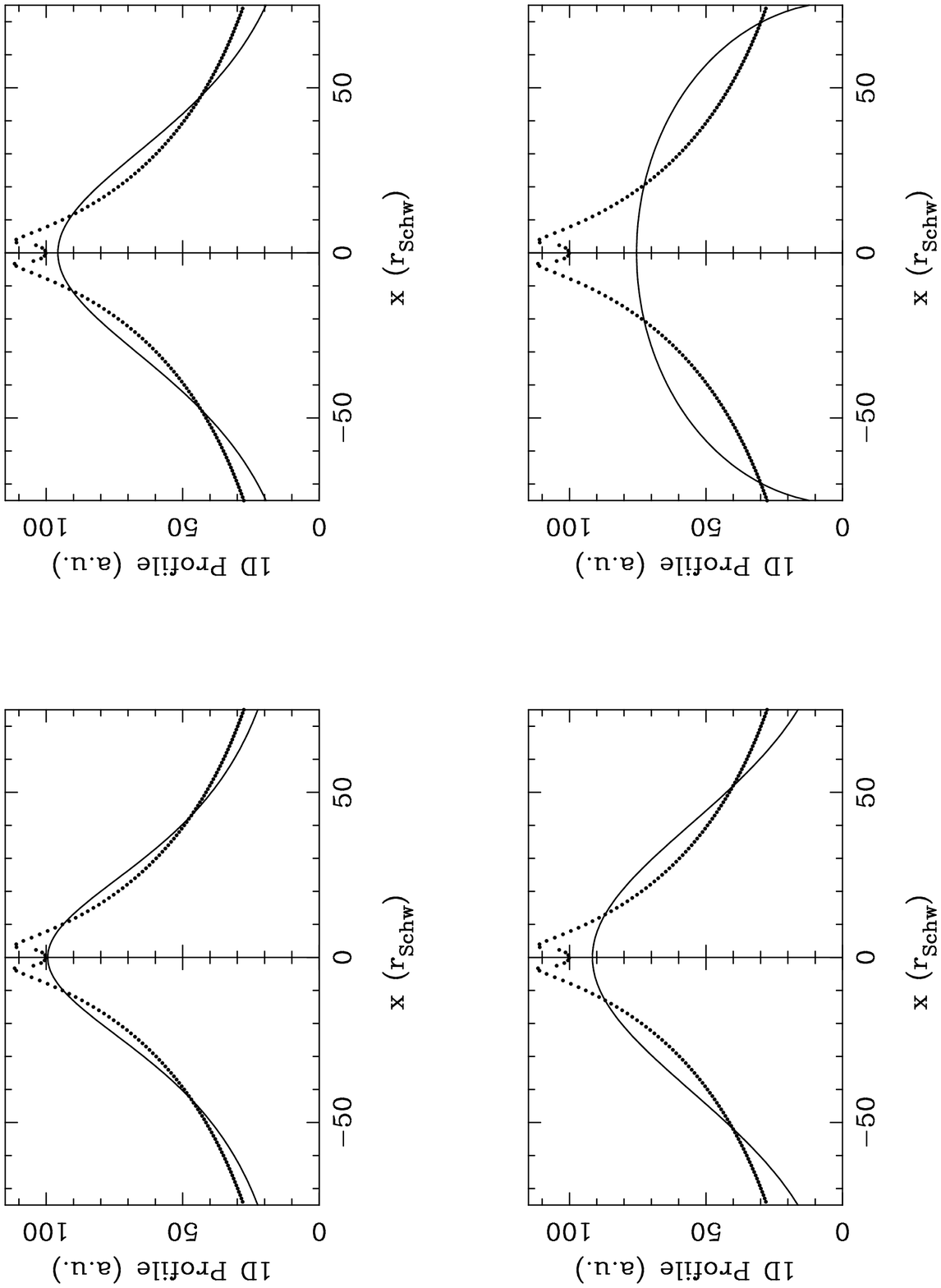]{Comparison of a given exact standard 1D intensity profile
(filled circles) and the effective 1D profiles fitting it (solid lines). The
effective source models are: $p$ = 3/2 power-law (left-hand top panel), $p$ = 
5/2 power-law (right-hand top panel), Gaussian (left-hand bottom panel), and 
top-hat (right-hand bottom panel).}

\setcounter{figure}{0}

\clearpage

\begin{figure}
\plotone{f1.eps}
\caption{}
\end{figure}

\clearpage

\begin{figure}
\plotone{f2t.eps}
\caption{(top)}
\end{figure}

\setcounter{figure}{1}

\clearpage

\begin{figure}
\plotone{f2b.eps}
\caption{(bottom)}
\end{figure}

\clearpage

\begin{figure}
\plotone{f3t.eps}
\caption{(top)}
\end{figure}

\setcounter{figure}{2}

\clearpage

\begin{figure}
\plotone{f3b.eps}
\caption{(bottom)}
\end{figure}

\clearpage

\begin{figure}
\plotone{f4.eps}
\caption{}
\end{figure}

\clearpage

\begin{table}
\begin{center}
\caption{Fits to the GLITP $V$-band light curve for Q2237+0305A.\label{tbl-1}}
\begin{tabular}{cccccc}
\tableline\tableline
Source model & $F_0$ (mJy) & $F_C$ (mJy) & $t_0$ (JD--2450000) & $\Delta t$
(days) & $\hat{\chi}^2 (min)/\delta (min)$ \\
\tableline
Uniform & 0.72 & 0.07 & 1489.6 & $30.2^{+1.3}_{-1.3}$ & 1.33/1.62 \\
Gaussian & 0.71 & 0.09 & 1488.6 & $31.0^{+8.3}_{-2.8}$ & 1.31/1.52 \\
Power-law ($p_V$ = 5/2) & 0.68 & 0.11 & 1486.1 & $48.0^{+12.1}_{-8.4}$ &
1.18/0.88 \\
Power-law ($p_V$ = 3/2) & 0.65 & 0.19 & 1484.7 & $33.1^{+8.3}_{-6.1}$ &
1.08/0.39 \\
Standard & 0.59 & 0.30 & 1480.6 & $39.6^{+10.4}_{-9.5}$ & 0.99/0.05 \\
\tableline
\end{tabular}
\end{center}
\end{table}

\clearpage

\begin{table}
\begin{center}
\caption{Fits to the GLITP $R$-band light curve for 
Q2237+0305A\tablenotemark{a}.\label{tbl-2}}
\begin{tabular}{cccccc}
\tableline\tableline
Source model & $F_0$ (mJy) & $F_C$ (mJy) & $t_0$ (JD--2450000) & $\Delta t$
(days) & $\hat{\chi}^2 (min)/\delta (min)$ \\
\tableline
Uniform & 1.28 & 0.09 & 1489.3 & $30.7^{+1.4}_{-1.1}$ & 1.44/2.09 \\
Gaussian & 1.18 & 0.17 & 1483.7 & $47.6^{+34.1}_{-9.0}$ & 1.30/1.42 \\
Power-law ($p_R$ = 5/2) & 1.14 & 0.20 & 1481.6 & $65.5^{+34.4}_{-18.0}$ &
1.19/0.90 \\
Power-law ($p_R$ = 3/2) & 1.12 & 0.33 & 1481.4 & $41.3^{+18.8}_{-10.9}$ &
1.12/0.57 \\
Standard & 1.06 & 0.46 & 1478.7 & $43.0^{+15.0}_{-11.3}$ & 0.99/0.05 \\
\tableline
\end{tabular}
\tablenotetext{a}{The observed flux at day 1473 deviates from neighbouring 
points and it was removed from our brightness record}
\end{center}
\end{table}

\clearpage

\begin{table}
\begin{center}
\caption{Source size ratios and bounds on the dimension of the $V$-band and
$R$-band sources.\label{tbl-3}}
\begin{tabular}{cccc}
\tableline\tableline
Source model & $q = R_V/R_R$ & $R_V(50\%) \times 10^2$ (pc) & 
$R_R(50\%) \times 10^2$ (pc) \\
\tableline
Uniform & $0.98^{+0.05}_{-0.06}$ & $<$ 0.044 (0.061)\tablenotemark{a} & 
$<$ 0.044 (0.063)\tablenotemark{b} \\
Gaussian & $0.65^{+0.21}_{-0.47}$ & $<$ 0.065 & $<$ 0.133 \\
Power-law ($p$ = 5/2) & $0.73^{+0.27}_{-0.40}$ & $<$ 0.091 & $<$ 0.150 \\
Power-law ($p$ = 3/2) & $0.80^{+0.29}_{-0.39}$ & $<$ 0.140 
(0.807)\tablenotemark{c} & $<$ 0.205 (1.172)\tablenotemark{d} \\ 
Standard & $0.92^{+0.34}_{-0.39}$ & $<$ 0.234 
(0.689)\tablenotemark{c} & $<$ 0.271 (0.800)\tablenotemark{d} \\
\tableline
\end{tabular}
\tablenotetext{a}{Upper limit on $R_V \times 10^2$ (pc)}
\tablenotetext{b}{Upper limit on $R_R \times 10^2$ (pc)}
\tablenotetext{c}{Upper limit on $R_V(90\%) \times 10^2$ (pc)}
\tablenotetext{d}{Upper limit on $R_R(90\%) \times 10^2$ (pc)}
\end{center}
\end{table}

\clearpage

\begin{table}
\begin{center}
\caption{Comparison between the {\it standard} fits to the GLITP and OGLE data.
\label{tbl-4}}
\begin{tabular}{cccccc}
\tableline\tableline
Dataset & $F_0$ (mJy) & $F_C$ (mJy) & $t_0$ (JD--2450000) & $\Delta t$
(days) & $\hat{\chi}^2 (min)$ \\
\tableline
GLITP & 0.59 & 0.30 & $1480.6^{+3.1}_{-3.2}$ & $39.6^{+10.4}_{-9.5}$ & 0.99 \\
OGLE99 & 0.72 & 0.20 & $1487.6^{+1.0}_{-1.0}$ & $14.3^{+3.4}_{-2.6}$ & 1.51 \\
OGLE99-00 & 0.60 & 0.34 & $1471.2^{+3.1}_{-2.8}$ & $29.5^{+3.2}_{-3.1}$ & 7.49 
\\
\tableline
\end{tabular}
\end{center}
\end{table}

\end{document}